\input{aipcheck}
\documentclass[
    ,final            
  ]
  {aipproc}

\layoutstyle{8x11double}

\begin{document}

%
%
%



\title{Deeply Virtual Neutrino Production of $\pi^0$ from Nucleon and Nuclear Targets}



\author{Gary R.~Goldstein}{ address={Physics Department,Tufts University,
		Medford, MA 02155} }
		
\author{Osvaldo Gonzalez Hernandez}{address={Physics Department, University of Virginia
		Charlottesville, VA 22904}  }

\author{Simonetta Liuti}{address={Physics Department, University of Virginia
		Charlottesville, VA 22904}  }
		
\author{Tracy McAskill}{address={Physics Department,Tufts University,
		Medford, MA 02155}  }

\begin{abstract}
In most experiments aimed at determining the masses and mixing parameters for neutrino oscillations
in different energy regimes there are strong backgrounds to neutrino interactions with matter that can confuse the expected small signals. 
One particular set of backgrounds is due to neutrino production of mesons, particularly pions, from nucleon or nuclear targets. It is especially important and 
theoretically interesting in itself to study $coherent$ neutrino production of pions, i.e. wherein the target does not break up. Preserving the target requires that the meson produced in the reaction will be charged for Charged Current (CC) interactions or neutral for Neutral Current (NC) interactions. Coherent scattering is the weak analog of exclusive neutral meson electroproduction for which considerable progress has been made in understanding the intermediate energy, small momentum transfer region by applying the Generalized Parton Distribution
(GPD)  phenomenology. The GPD perspective will be explained with emphasis on how the understanding gained from electroproduction of neutral pions can carry over to the neutrino processes. 
\end{abstract}
\maketitle

\begin{figure}
  \includegraphics[height=0.2\textheight]{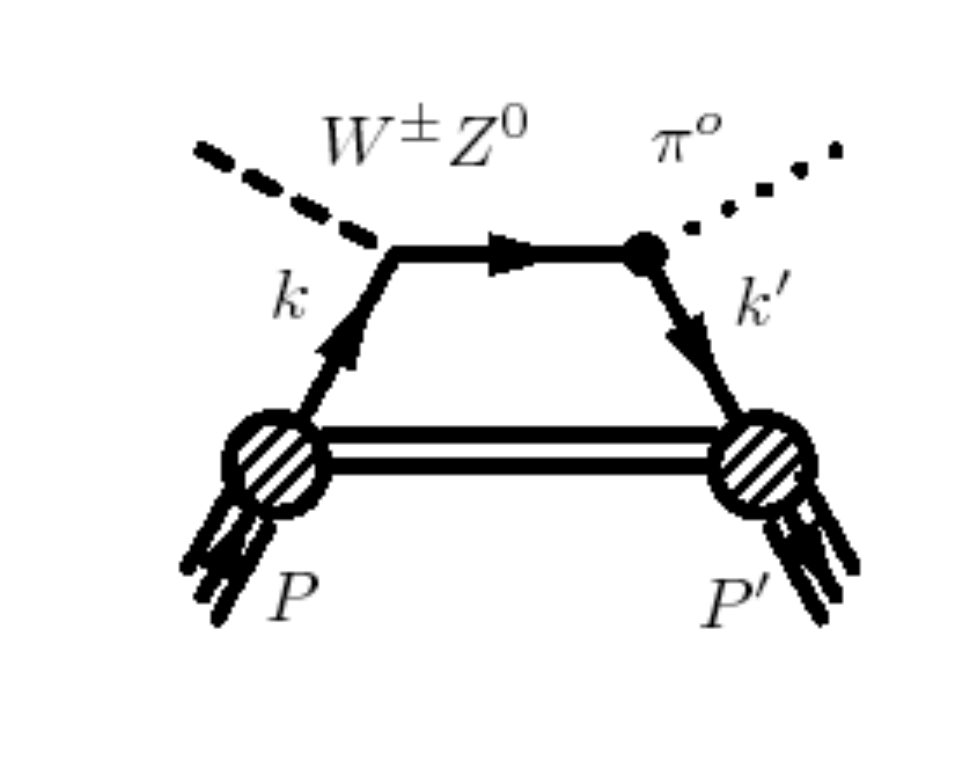}
  \caption{Diagram for $\nu P \rightarrow \pi^o P^\prime$ process.}
  \label{handbag}
\end{figure}

Generalized Parton Distributions (GPDs) have been employed very fruitfully in describing Deeply Virtual Compton Scattering (DVCS) and the exclusive electroproduction of mesons (see Refs.\cite{Mueller,Ji,Rad} for reviews). 
The essential physics in these processes is that highly virtual photons carry a hard scale, $Q^2$, but interact with the partons within the nucleon, whose momentum distribution within the nucleon is characterized by a soft scale, $ \approx \Lambda_{QCD}$. 
At leading order for DIS, this is exemplified by the well known ``handbag'' diagram.  
Since the processes that we consider here, however, are {\em exclusive}, the handbag represents a "non-forward" scattering amplitude where the parton rejoins the outgoing nucleon with different 4-momentum from the target nucleon (Fig. \ref{handbag}). 
To leading order in the hard momenta, the amplitude for the process factorizes into two parts. 
One factor is the pQCD calculable 
$\gamma^*$+parton$\rightarrow$ meson or $\gamma + parton^\prime$ - upper part of the handbag. 
The soft factor, the amplitude (or quark or gluon field correlator) for quark or gluon depends on the change of the nucleon 4-momentum as well as the 4-momentum of the parton relative to the nucleon. This soft part, the lower part of the handbag describes the GPDs. Specifically,
in the DIS limit, in addition to the parton Light Cone (LC) momentum fraction, $X=k^+/P^+$, and the virtual boson's four-momentum, $Q^2$, one now introduces two additional variables, the so-called skewness,
$\zeta$, defined through the final quark LC momentum fraction, $X-\zeta=k^{\prime +}/P^+$, and the
four-momentum transfer squared between the initial and final nucleons, $t=(P-P^\prime)^2$ (see Fig.\ref{handbag}). By working out the kinematics in the Bjorken limit, one finds out that $\zeta \approx x_{Bj}=Q^2/2M_N \nu$, $\nu$ being the virtual boson's energy. Furthermore, $X$ is integrated over in the evaluation of the amplitude. The observables that we define below can be therefore expressed as functions of the invariants: $x_{Bj} \approx \zeta$, $Q^2$, and $t$.   
 
Taking the spins of nucleon and the quarks (as the partons) into account leads to eight distinct GPDs for different combinations 
of helicities~\cite{HooJi,Die_01}.
In the following we apply this factorized picture to neutrino production of pions from nucleons or nuclei. 
The same GPDs that are involved in electroproduction are involved here, as we will show.  Note that parity violation acts on the upper, hard part of the handbag, but not the lower GPDs.

The GPD based analysis is similar to what we developed for electroproduction in Ref. ~\cite{AhmGolLiu} (proton targets) and Ref.~\cite{LiuTan,GolLiu} (spin 0 nuclear targets). 
The spin dependences and the question of the extraction of observables from experiments will be addressed in some detail in this presentation. Before considering the formulation of the GPD framework, let us recall the salient features of the phenomenology. We will deal with helicity amplitudes throughout, particularly for the virtual weak boson scattering from a nucleon to produce a meson and the scattered nucleon. There are two planes in this process, the lepton scattering plane and the ``hadron'' plane. We are most interested in the dependence on the azimuthal angle between these planes.  

The differential cross section is
 \begin{eqnarray}
 \frac{d^4 \sigma }{d\Omega d\epsilon _2 d\phi dt} & = & \Gamma \left[ \displaystyle \frac{d\sigma _T }{dt} + \epsilon _L \frac{d\sigma _L }{dt} 
 + \epsilon \cos 2\phi \frac{d\sigma _{TT}}{dt}   \right.
 \nonumber \\   
&  + & \sqrt {2\epsilon _L (\epsilon  + 1)}   \cos \phi \frac{d\sigma _{LT} }{dt} 
\nonumber \\   
& +  & \epsilon \sin2\phi \frac{d\sigma _{T'T}}{dt}  
\nonumber \\  
&  \pm  & \left. \sqrt {2\epsilon _L (\epsilon  + 1)}  \sin \phi \frac{d\sigma _{L'T} }{dt} \right] 
\label{xsection}  
\end{eqnarray}
It can be seen that there are two terms here (that do not appear in electroproduction) that involve parity violation, the $\sin \phi$ and $\sin 2\phi$ cross sections. They arise from the interference between the Vector (V) and Axial Vector (A) weak currents. 
The $\sin \phi d\sigma_{L^\prime T}$ involves longitudinal vector amplitudes interfering with transverse axial vector amplitudes, and {\it vice versa}. 
This kind of interference is obtained for electroproduction by having a polarized electron beam, whereas for neutrino production the beam is always polarized (ignoring the very small mass effects). 
It produces an asymmetry between $\pi$'s produced above the lepton scattering plane and those below.
The $\sin2\phi d\sigma _{T'T}$ term involves transverse vector amplitudes interfering with transverse axial Vector amplitudes. This gives rise to an asymmetry between $\pi$s produced in the first and third azimuthal quadrant relative to the lepton plane and the second and fourth quadrants. Hence these interference effects can be obtained by carefully measuring the azimuthal dependence. We will see that these effects reveal important structure of the soft process through the GPDs.  

These partial cross sections are bilinears in the helicity amplitudes, of which there are 12. There is a doubling of the number of independent helicity amplitudes, 
$f_{\Lambda, \lambda_N; 0, \lambda_{N^\prime}}$ from 6 in electroproduction to 12 in the neutrino case. 
There is no longer an equality, up to a sign, between $f_{\Lambda,\lambda_N;0,\lambda_{N^\prime}}$ and $f_{-\Lambda,-\lambda_N;0,-\lambda_{N^\prime}}$. However, by separating the V and A contributions, we see that one involves the positive and the other the negative phase between parity reflected helicity amplitudes. 

In the handbag approximation that will be adopted here, the helicity amplitudes (the $f_{\Lambda, . . .}$'s) are given by a convolution of the upper, hard amplitude, 
$g_{\Lambda,\lambda; 0,\lambda^\prime}$, and the lower, soft part, the GPD helicity amplitudes, $A_{\lambda_N,\lambda; \lambda_{N^\prime}, \lambda^\prime}$. The convolution involves summing over intermediate quark helicities and integrating over unobservable quark momenta, constrained by the external momenta
\begin{equation}
f_{\Lambda,\lambda_N; 0, \lambda_{N^\prime}}=\sum_{\lambda,\lambda^\prime} g_{\Lambda,\lambda;0,\lambda^\prime} \otimes A_{\lambda_N,\lambda;\lambda_{N^\prime},\lambda^\prime}.
\label{hel_amp}
\end{equation}
The helicity amplitudes $A_{\Lambda, . . .}$ involve some or all of the 8 GPDs (always doubled by the distinction between crossing even or odd combinations), for example  
\begin{eqnarray}
A_{++,--} & \rightarrow & \sqrt{1-\xi^2} \left[ H_ T + \frac{t_0-t}{4M^2} \widetilde{H}_T \right.
\nonumber \\
 & - & \left. \frac{\xi^2}{1-\xi^2}E_T + 
\frac{\xi}{1-\xi^2}\widetilde{E}_T \right] 
\label{Aodd}
\end{eqnarray}
\begin{equation}
A_{++,++} \rightarrow \frac{1}{2}\sqrt{1-\xi^2} \left[ H + \widetilde{H} - \frac{\xi^2}{1-\xi^2}(E+\widetilde{E})\right].
\label{Aeven}
\end{equation}
We have suppressed the variables $X, \zeta, t, Q^2$, and we have omitted the integration over $X$  for brevity. The first amplitude involves the chiral odd GPDs, for which $H_T$, in particular, is connected to the {\it tensor charge} and transversity of the nucleon. The second involves the chiral even GPDs, measured indirectly in DVCS. We will see that our preferred model for $\pi^0$ production involves the chiral odd GPDs and thereby gives us a window into transversity.

We consider the general structure of the weak amplitudes, the hadronic tensor, by analogy with electroproduction. For electroproduction of an on-shell photon, or DVCS, the amplitudes are given by a Fourier transform of the correlation function
of two electromagnetic currents. With currents ($J_{EM}^\mu$ and
$J_{EM}^\nu$), the amplitude can be written as
\begin{eqnarray}
T_{EM}^{\mu\nu} & = & i
\int d^{4}z \; e^{-i\bar{q}\cdot z}  \nonumber \\ & & 
\left\langle P^\prime, S^\prime  \right| 
T \left\{ J_{EM}^{\mu} (z/2) J_{EM}^{\nu}(-z/2 )\right\} 
\left| P, S \right\rangle\ .
\label{eq:standardamplitude}
\end{eqnarray}
where $p_1$ and $p_2$ are the four-momenta of the initial and final
nucleons, $s_1$ and $s_2$ are their spins, and $\bar{q}=(q_1+q_2)/2$ is the average of the incoming and outgoing boson 4-momenta. In this case the connection with the helicity amplitudes is given by 
\begin{eqnarray}
f_{\Lambda, \lambda_N; \Lambda^\prime, \lambda_{N^\prime} } 
= \epsilon^{\prime * \, \Lambda^\prime} _\mu \, T_{EM \, (\lambda_N, \lambda_{N^\prime})}^{\mu\nu} \, \epsilon_\nu^{\Lambda} 
\label{hel_amp_EM}
\end{eqnarray}
where we have written the polarization indices explicitly. 
For leptoproduction of mesons, the leftmost current is replaced by the hadronic current for the meson.
In particular, the weak $\pi^0$ production process, with an incoming $Z^0$ or $W^\pm$
boson and outgoing pion, has the amplitude
\begin{eqnarray}
T_{W}^{\mu} & = & i\int d^{4}z \; e^{-i\bar{q}\cdot z}  \nonumber \\ & & 
\left\langle P^\prime, S^\prime  \right| 
T \left\{ J_{\pi^0} (z/2) J_{W}^{\mu}(-z/2 )\right\} 
\left| P, S \right\rangle\ .
\label{eq:weakComptonamplitude1}
\end{eqnarray}
$J_{W}^{\mu}\left(z/2\right)$ is either the weak
neutral current $J_{WN}^{\mu}\left(z/2\right)$, or the weak charged
current $J_{WC}^{\mu}\left(z/2\right)$. 
For $\pi, \eta, \rho, \omega$, etc., the GPDs can carry isospin change. For $K$ production there will be a hyperon recoiling, requiring a strangeness changing GPD. 

To leading order in $1/Q$ the connection with the helicity amplitude given in Eq.(\ref{hel_amp}) reads
\begin{eqnarray}
f_{\Lambda, \lambda_N; 0, \lambda_{N^\prime} } 
& = & \sqrt{2} G_F  \left[ \bar{u}(k^\prime)\gamma_\mu \left(c_V-\gamma_5 c_A \right) u(k) \right]_\Lambda \,
\nonumber \\
& \times &  T_{W \, (\lambda_N, \lambda_{N^\prime})}^{\mu} ,
\label{hel_amp_nu}
\end{eqnarray}
where the term containing $u(k^\prime)$ and $u(k)$, the spinors for the outgoing and incoming {\it leptons}, respectively, can be also written in terms of
the  Weak Vector Boson polarization vector. 

 
Note that the $\gamma^\mu \gamma^5$ coupling leads to a form similar to the photon production case ($\nu DVCS$) explored by Psaker  {\it et al.}~\cite{PsaMelRad}, with vector and axial vector couplings switched, 
%
\begin{equation}
\sum_{f} \left[s^{\mu\rho\nu\eta}\left( c_{A}^{f} \mathcal{G}_{\eta}^{-} - c_{V}^{f}\mathcal{G}_{5\eta}^{-}\right) 
-i \epsilon^{\mu\rho\nu\eta} \left( c_{A}^{f} \mathcal{G}_{5\eta}^{+} -  c_{V}^{f} \mathcal{G}_{\eta}^{+}\right) \right]   
\label{eq:weakneutralexpansion2}
\end{equation}
where
\begin{eqnarray}
\label{eq:uncontractedstringoperators1}
\mathcal{G}_{\Gamma}^{f\pm} & \equiv &
\left[\bar{\psi}_{f}\left(z/2\right)\Gamma\psi_{f}
\left(-z/2\right)\pm\left(z\rightarrow-z\right)\right], 
\label{eq:uncontractedstringoperators2}
\end{eqnarray}
$\Gamma$ being a Dirac matrix
and $s^{\mu\rho\nu\eta}\equiv g^{\mu\rho}g^{\nu\eta}+
g^{\mu\eta}g^{\rho\nu}-g^{\mu\nu}g^{\rho\eta}$. 

As we emphasized in Ref.\cite{AhmGolLiu}, $\pi^0$ production also involves a  C-Parity odd and chiral odd diagram in the $t-$channel. 
Therefore, the corresponding Dirac structure for the hard subprocess diagram involves $\sigma^{+T}\gamma^5$, 

One can now isolate the different types of Lorentz structures involving tensor, vector and axial vector components in the  nucleon matrix elements of the operators in Eq.(\ref{eq:weakComptonamplitude1}) 
\cite{neutrinopaper}.

For the tensor operators the expressions are given by linear combinations involving the four chiral odd GPDs~\cite{Die_01}
\begin{eqnarray}
 & \frac{1}{2 P^+}  \, \overline{U}(P^\prime, S^\prime) \,
 [ \mathcal{H}_T^q \, i \sigma^{+ \, i} +  \mathcal{\widetilde{H}}_T^q  \,  \displaystyle\frac{P^+ \Delta^i - \Delta^+ P^i}{M^2}  \, & \nonumber \\
 & +\mathcal{E}_T^q \,  \displaystyle\frac{\gamma^+ \Delta^i - \Delta^+ \gamma^i}{2M} \,
  + \mathcal{\widetilde{E}}_T^q \, \displaystyle \frac{\gamma^+P^i - P^+ \gamma^i}{M}]  U(P,S) &
\label{oddgpd} 
\end{eqnarray}
where 
$q=u,d,s$. 
For the vector term two chiral even GPDs are involved,
\begin{equation}
 \frac{1}{2 P^+}  \, \overline{U}(P^\prime, S^\prime) \,
 [ \mathcal{H}^q \, \gamma^+ 
 +\mathcal{E}^q \, i\frac{\sigma^{+\alpha} \Delta_\alpha}{2M} ], 
 U(P,S) 
\label{vectorgpd} 
\end{equation}
and for the axial vector 
\begin{equation}
 \frac{1}{2 P^+}  \, \overline{U}(P^\prime, S^\prime) \,
 [ 
\mathcal{\widetilde{H}}^q \gamma^+ \gamma^5 +
\mathcal{\widetilde{E}}^q \frac{\gamma^5\Delta^+}{2M}]
 U(P,S) .
\label{axialgpd}
\end{equation}
The resulting functions, $\mathcal{H}(\xi,t), \mathcal{H}_T(\xi,t)$, ..., that enter the helicity amplitudes $f_{\Lambda, . . . }$ 
are complex. They are the
 analogs of the Compton form factors in DVCS.  The quark's unobserved fractional longitudinal momentum $X$ appearing in  the GPDs is
integrated over as
\begin{eqnarray}
\mathcal{F}^q(\zeta,t)  = i \pi 
\left[ F^q(\zeta,\zeta,t) - F^{\bar{q}}(\zeta,\zeta,t) \right] +
\nonumber \\ 
\mathcal{P} \int\limits_{-1+\zeta}^1 dX \left(\frac{1}{X-\zeta} + \frac{1}{X} \right) F^q(X,\zeta,t)
\label{cal_F}
\end{eqnarray}
where $\mathcal{P}$ indicates a principal value integration and 
$\mathcal{F}^q = \mathcal{H}_T^q, \mathcal{E}_T^q, \widetilde{\mathcal{H}}_T^q, \widetilde{\mathcal{E}}_T^q$ (similarly for the chiral even sector).



The helicity amplitudes result from  contracting the various Lorentz structures described above with the polarization vector and the nucleon spinors evaluated for each combination. In Ref.~\cite{AhmGolLiu}, for the electroproduction of $\pi^0$s, this was carried out as we indicated in Eq.~\ref{Aodd} and Eq.~\ref{Aeven}.
For neutrino-production the structure of the kinematic and Dirac matrices is complicated by the parity odd parts. 
This is not a major problem, however, since it is taken care of by the hard scattering process at the upper part of the handbag, as noted
(more details will be given in Ref.\cite{neutrinopaper}).  
 
The general form of the cross section observables in terms of helicity bilinears is given in Ref.~\cite{AhmGolLiu}. The $\sin \phi$ term is of special interest and is given by 
\begin{eqnarray}
\frac{d\sigma_{L^\prime T}}{dt} & = & 2 \, \mathcal{N} \,  
\Im m [ f_{0,+;0,+}^{V*} (f_{1,+;0,-}^A + f_{1,-;0,+}^A) \nonumber \\ 
 & + & f_{0,+;0,-}^{V*} (f_{1,+;0,+}^A - f_{1,-;0,-}^A) 
 \nonumber \\
 & + & (V \leftrightarrow A) ]
\label{dsigLTp}
\end{eqnarray}
involving the longitudinal and transverse parity violating interference. Note that $ f_{1,+;0,-}$ will involve $H_T$, even in the small $t$ limit, so the tensor charge will be probed by this $\sin \phi$ term. 

To see how this cross section can be realized, we use the parametrization for GPDs that was applied in the electroproduction of $\pi^0$'s. The result for this model is shown in Fig. 2. Sizable interference is predicted, as was the case in predicting the beam asymmetry for electroproduction in the Jefferson Lab kinematic regime. Furthermore, a Regge model for the same asymmetry in the neutrino case shows a spectacular behavior. The asymmetry peaks at 15\% for $t= -0.4$. For electroproduction that asymmetry requires the beam spin asymmetry be measured. For neutrinos it is the $F_3$ type term and is measured in the azimuthal asymmetry directly. If we stick to ``factorization'' this will be zero. It is very important to see this as non-zero. It requires both longitudinal and transverse Weak boson coupling. 
Hence it is linear in the transversity of the target, providing another method for determining this elusive quantity. 
\begin{figure}
  \includegraphics[height=0.35\textheight]{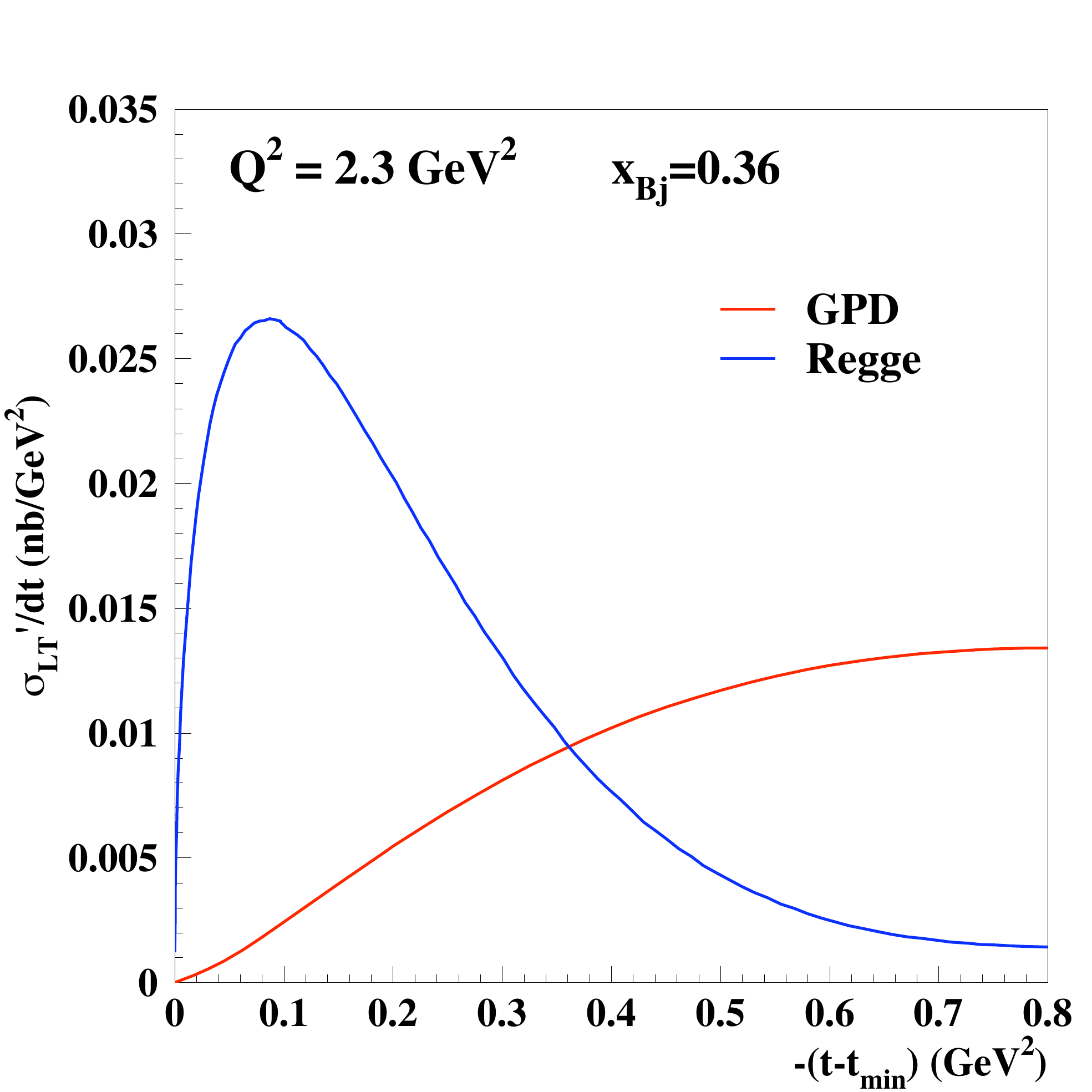}
  \caption{The contribution to the $\pi^o$ exclusive $\nu$ production cross section, $d \sigma_{L ^\prime T}/dt$, in both the GPD and Regge descriptions in the few GeV 
  region accessible in future neutrino experiments.}
\end{figure}

The other observables are readily calculated and will be presented in a forthcoming paper \cite{neutrinopaper}. How are these predictions to be tested?
How are experiments to see $\phi$ dependence? Charge Current interactions like $\nu+N\rightarrow \mu^- \pi^+ + N$  allows 2 planes to be seen. The GPDs probed are diagonal and have been modeled. 
What is to be expected? From $e+N\rightarrow e \pi^0 + N$ it is clear that $d\sigma_{L^\prime T} /dt $, shows the importance of transverse photons. Our calculations for neutrino processes indicate similar enhancements.  

Regarding CC that changes the quark flavor, (u into d or reverse), as in ${\bar \nu}+p\rightarrow \mu^+ \pi^0 + n$, the GPDs are non-diagonal, but simply related by isospin to the diagonal terms. However, care must be taken in applying the GPDs decomposition directly because of the importance of the charged pion pole exchange for the small $|t|$ region. The inclusion of these exchanges is being studied and related to the axial coupling through PCAC, as well known in the low energy regime.

Finally, we comment on neutrino scattering from nuclear targets in the multi-GeV/DIS region \cite{GolLiu}. An accurate study of nuclear effects is fundamental even in the DIS region where they have been found to be considerable in a consistent set of measurements carried out throughout the years. 
Nuclear GPDs have just begun to be explored in electron  scattering experiments 
at both Hermes \cite{Hermes} and Jefferson Lab \cite{Hafidi}. One hopes that GPDs will provide new insights on the elusive EMC effect problem, namely the apparent modification of the short distance scale structure of nuclei with respect to the nucleon's one.  Although the EMC effect can be considered as an indication of the modification of quarks and gluons
distributions inside nuclei, a unifying picture of the various explanations put forth to explain the effect is still lacking. Measuring GPDs provides a new framework where both
the intrinsic motion of quarks and gluons and their spatial distribution can in principle be pinned down.  Most nuclear targets of interest in neutrino experiments 
have spin zero.  This is a desirable feature because the number of nuclear helicity amplitudes and therefore of GPDs reduces to two, in the chiral even and chiral odd sectors, respectively, thus making it simpler to extract the nuclear effects. 

The amplitude for neutral pion production is written in terms of the chiral odd amplitude as
\begin{equation}
f_{\Lambda_\gamma,0; 0, 0} =  \sum_{\lambda,\lambda^\prime} 
g_{\Lambda_\gamma,\lambda;0,\lambda^\prime}  \otimes
C_{0,\lambda^\prime;0,\lambda}.
\end{equation}
$\Lambda_\gamma = \pm 1,0$ being the virtual photon spin.
The $C_{0,\lambda^\prime;0,\lambda}$ are the ``quark-nucleus'' helicity amplitudes. 
that depend on $x_{Bj}, t$ and $Q^2$ while 
implicitly containing an integration over unobserved quark and nucleon momenta. 
They can be written in terms of the quark-nucleon helicity amplitudes:
\begin{equation}
C_{0,\lambda^\prime;0,\lambda} = \sum_{\Lambda_N, \Lambda_N^\prime} \int d^4 \, P \, B_{0,\Lambda_N^\prime;0,\Lambda_N} 
A_{\Lambda_N^\prime\lambda^\prime;\Lambda_N, \lambda^\prime}
\end{equation} 
Two terms survive
\begin{eqnarray}
T \; \; \; & \Rightarrow  & \; \; \; g_{1+,0-} \, C_{0-;0+}  \\
L \; \; \; & \Rightarrow  & \; \; \; g_{0+,0-} \, C_{0-;0+}  
\end{eqnarray}
Both terms contain the same $C$ function.
The latter is given by
\begin{eqnarray}
C_{0,-;0,+} & = & \int d^4 P \left[ B_{0+.0-} A_{+-;-+} +
 B_{0-;0-} A_{--;-+} + \right. \nonumber \\
& &  \left. B_{0+;0+} A_{+-;-+} +
 B_{0+.0+} A_{+-;-+} \right], 
\end{eqnarray}
where 
\[ B_{0+;0+}  =   B_{0-.0-} = \cos \frac{\theta}{2} \rho_A(P^2,P^{\prime \, 2}), \]
and 
\[ B_{0+;0-}  =  - B_{0-.0+}  \sin \frac{\theta}{2} \rho_A(P^2,P^{\prime \, 2}). \]
Combining these equations with the results for the nucleon GPDs \cite{AhmGolLiu} one obtains
\begin{eqnarray}
C_{0,-;0,+} & =  & F_V(Q^2) \int d^4 P \, \rho_A(P^2,P^{\prime \, 2}) 
\nonumber \\
& \times & \left[ \sin \frac{\theta}{2} \left(f_3-f_2 \right) +\cos \frac{\theta}{2} \left(f_1+f_4 \right) \right]
\end{eqnarray} 
where $f_1 = f_4 = A_{++,+-}$, $f_2 = A_{--,++}$, $f_3 =  A_{++,--}$, are the same ones appearing 
in the nucleon case \cite{AhmGolLiu}, but 
evaluated at off-shell kinematics.  $F_V$ is the vector contribution to the form factor describing the neutral pion
vertex.
The nucleons spectral function is
\begin{eqnarray}
\rho_A(P^2,P^{\prime \, 2})  & =  & \sum_f \Phi_f(\mid {\bf P} \mid ) \Phi_f^*(\mid {\bf P^\prime} \mid)
\nonumber \\
& \times & \delta\left( E-(E_{A-1}^f -E_A) \right). 
\label{rho_A}
\end{eqnarray} 
where $E_A$ is the nuclear binding and $E_{A-1}^f$ is the nucleon removal energy.
A considerable effect was predicted within a class of models in Ref.\cite{LiuTan}, shown in Fig.3. 
\begin{figure}
  \includegraphics[height=0.35\textheight]{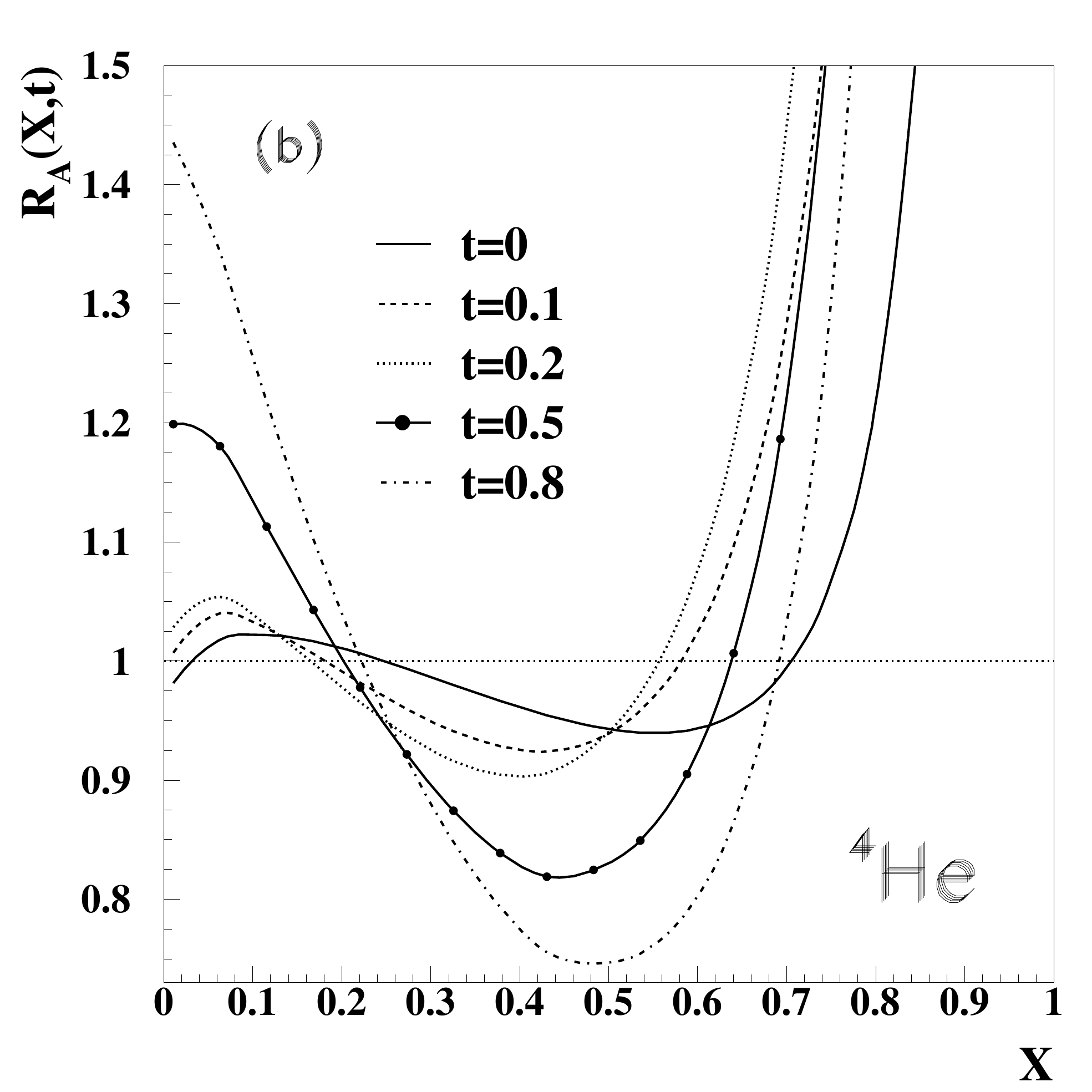}
  \caption{The Generalized EMC effect in $^4He$.  Increasingly larger nuclear
  effects with increasing momentum transfer squared, $t$, are predicted.(Adapted from Ref.\cite{LiuTan})}
\end{figure}
These calculations are being currently extended to
neutrino scattering experiments \cite{neutrinopaper}.

In conclusion, we have shown how the intermediate $Q^2$ kinematics of coherent neutrino production of pions can be usefully interpreted in the GPD framework. 
This will enable the insights gained from electroproduction to be extended to the neutrino processes.

\begin{theacknowledgments}
G.G. and S.L. thank the organizers of NuFact09 for inviting us. We appreciate discussions with H. Gallagher. S.L. thanks Kawtar Hafidi and Roy Holt  for their hospitality at Argonne National Lab where part of this work was completed. This work is supported by the U.S. Department of Energy grants no. DE-FG02-01ER4120 (O.G-H. and S.L), and 
no. DE-FG02-92ER40702 (G.R.G. and T.M.).
\end{theacknowledgments}

\end{document}